\documentclass[prb,twocolumn,floatfix,showpacs,superscriptaddress]{revtex4}
\usepackage{inputenc}
\usepackage{graphicx,epsfig,amssymb}

\begin{document}

\title{Range separated hybrid density functional study of organic dye 
sensitizers on anatase TiO$_2$ nanowires}

\author{Hatice \"{U}nal}
\affiliation{Department of Physics, Bal{\i}kesir University, 
Bal{\i}kesir 10145, Turkey}

\author{Deniz Gunceler}
\affiliation{Department of Physics, Cornell University, 
Ithaca, NY 14853, USA}

\author{O\u{g}uz G\"{u}lseren}
\affiliation{Department of Physics, Bilkent University,
Ankara 06800, Turkey}

\author{\c{S}inasi Ellialt{\i}o\u{g}lu}
\affiliation{Basic Sciences, TED University, Ankara 06420, Turkey}

\author{Ersen Mete}
\email{emete@balikesir.edu.tr}
\thanks{Corresponding author}
\affiliation{Department of Physics, Bal{\i}kesir University, 
Bal{\i}kesir 10145, Turkey}

\date{\today}

\begin{abstract}
The adsorption of organic molecules coumarin and the donor-$\pi$-acceptor type 
tetrahydroquinoline (C2-1) on anatase (101) and (001) nanowires have been
investigated using screened Coulomb hybrid density functional theory
calculations. While coumarin forms single bond with the nanowire surface,
C2-1 additionally exhibits bidentate mode giving rise to much stronger
adsorption energies. Nonlinear solvation effects on the binding
characteristics of the dye chromophores on the nanowire facets have 
also been examined. These two dye sensitizers show different electronic charge 
distributions for the highest occupied and the lowest unoccupied molecular
states. We studied the electronic structures in terms of the positions of the
band edges and adsorbate related band gap states and their effect on the
absorption spectra of the dye-nanowire combined systems. These findings were
interpreted and discussed from the view point of  better light harvesting and
charge separation as well as in relation to more efficient charge carrier
injection into the semiconductor nanowire.
\end{abstract}


\maketitle

\section{Introduction}

Recently, dye sensitized solar cells (DSSC) have become an important research 
field in direct energy generation from sunlight. In a typical system, dye 
adsorbates on TiO$_2$ nanostructures are responsible for light harvesting with
their highest occupied molecular orbitals (HOMO) residing in the band gap of 
the host semiconductor. The coupling of their lowest unoccupied molecular
orbital (LUMO) with the conduction band (CB) of TiO$_2$ serves as a natural
pathway for photo-generated charge injection from the dye to the CB of the
substrate. The cell is then regenerated by interaction of the excited dye
with a redox couple.\cite{ORegan,Fujishima1} The open-circuit voltage of the
cell, V$_{\tiny\textrm{OC}}$,
is the difference between the highest occupied level of dye-semiconductor 
system and redox potential of the mediator (typically iodide). For such a solar
cell construction, the band positions of titania is one of the most appropriate 
among the other wide-bandgap semiconductors. 

TiO$_2$, particularly the (001) surface of anatase phase, show excellent 
photocatalytic activity under UV irradiation.\cite{Fujishima1,Fujishima2}
Sensitizer chromophores not only drive the UV-limited photoresponse of
TiO$_2$, into the visible range but also play an important role in charge 
carrier dynamics. In fact, the overall cell efficiency depends on the 
preferable properties of dye-semiconductor composite system in relation to the 
factors such as the photoelectric conversion, charge carrier injection, 
electron-hole recombination rates and charge transport performance.

High surface-to-volume ratio of the semiconductor material is another 
component of the efficiency consideration. In general, TiO$_2$ nanoparticles 
provide multiple surfaces exposing large number of active sites.
On the other hand, quasi-one-dimensional titania nanowires~\cite{Cakir-wire}
not only 
accommodate even larger areas but also are superior in n-type conductivity
reducing the photogenerated charge recombination rates. In the forms of 
nanostructures, thermodynamically the most stable phase of TiO$_2$ is the 
anatase polymorph.\cite{Naicker,Iacomino,Fuertes} The interaction between 
anatase nanostructures and dye sensitizers is one of the basic issues to 
improve the efficiency of DSSCs. 

The choice of the sensitizer dye depends not only on its durability, 
absorption and charge injection ability during the energy conversion 
cycles, but also on its production cost and easiness. Ru based 
photosensitizers, reaching up to 11$\%$ efficiencies (under AM 1.5 
illumination), constitute an important part of the dye complexes that 
are used in DSSC technology.\cite{ORegan,Nazeeruddin1,Tachibana,Thompson,Nakade,Wang1,Benko,Wang2,Nazeeruddin2} 
Research has been focused on finding alternatives to those well-known, 
metal-driven and relatively more expensive dye sensitizers.\cite{Mishra}
Several organic dye molecules have been shown to be strikingly efficient 
in light harvesting.\cite{Khazraji,Sayama,Hara1,Wang3,Qin,Koumura,Jiang,Chen1,Chen2} 
For instance, indoline derivatives showed 9.52$\%$ efficiency.\cite{Ito}
High efficiencies of the Ru complexes are attributed to their high charge 
injection rates into the conduction band (CB) of the TiO$_2$ pertaining to 
their metal-to-ligand charge transfer ability. Novel organic donor-$\pi$-acceptor 
dyes like tetrahydroquinoline based C2-1\cite{Chen1,Chen2} have been 
proposed to achieve intramolecular charge separation.

\begin{table*}[bth]
\caption{Calculated adsorption energies of dye-nanowire systems (in eV)\label{table1}}
\begin{ruledtabular}
\begin{tabular}{ccccc|cccc}
& \multicolumn{4}{c}{@(001)} &
\multicolumn{4}{c}{@(101)}\\ \cline{2-9}
Dye & PBE & HSE & PBE+PCM\footnote{Nonlinear PCM included for CHCl$_3$}& PBE+PCM\footnote{Nonlinear PCM included for H$_2$O}
& PBE & HSE & PBE+PCM$^a$ & PBE+PCM $^b$\\[1mm] \hline
Coumarin & $-0.46$ & $-0.63$ & $-0.24$ & $-0.05$& $-0.63$ & $-0.70$ & $-0.48$ &$-0.23$\\ \hline
C2-1(monodentate) & $-0.72$ &$-0.73$ & $-0.71$ &$-0.22$& $-0.57$ & $-0.62$ & $-0.44$ &$-0.10$\\ \hline
C2-1(bidentate) &$-1.36$ & $-1.25$ & $-1.22$ &$-0.71$& $-0.94$ & $-0.83$ & $-0.74$ &$-0.37$ \\ 
\end{tabular}
\end{ruledtabular}
\end{table*}

\begin{figure}[b!]
\epsfig{file=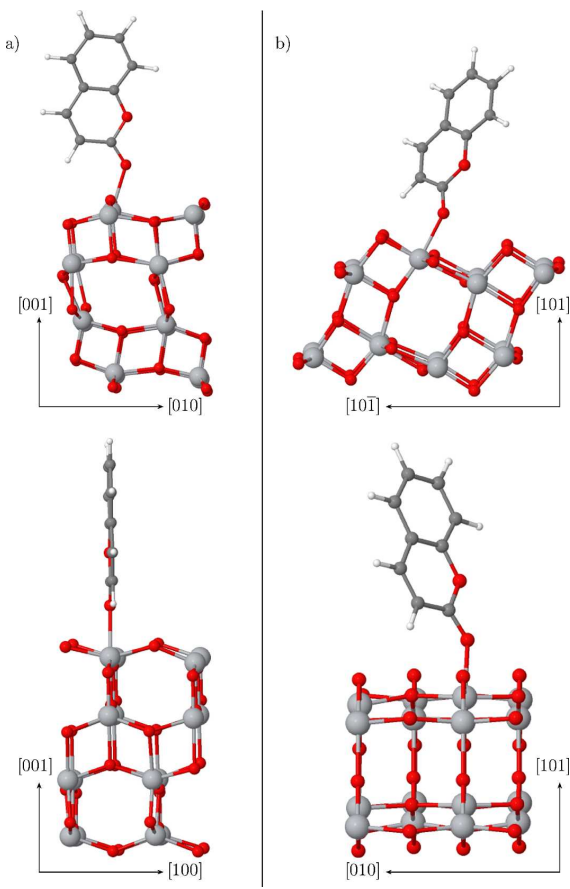,width=7.5cm}
\caption{Optimized adsorption geometries of coumarin on anatase (001)-nanowire 
(left panel) and (101)-nanowire (right panel) shown from two different viewing 
orientations. Here the red, light-grey, dark-grey, and white color spheres 
represent O, Ti, C, and H atoms, respectively.\label{fig1}}
\end{figure} 

Improvements in the dye design crucially depend on the deep understanding 
of the fundamental properties of the sensitizer and its interaction with 
the nanostructured semiconductor. Quantitatively, novel organic dye 
molecules with advanced chemical and physical properties can be tailored 
with the aid of theoretical modeling.\cite{Cakir1,Cakir2} Theoretical 
studies have focused on the prediction of the electronic structure of 
the chromophores and of their interface with the TiO$_2$ 
substrates.\cite{Daul,Brocks,Naito,Gorelsky,Stier,Acebal,Pasveer,Hara2,Prieto,Campbell,Persson}
In particular, the family of  tetrahydroquinoline derived dyes have been 
studied with \textit{ab initio} calculations.\cite{Zhang2}.  O'Rourke 
\textit{et al} have also considered these dyes on TiO$_2$(101) surface to 
predict their electronic structures.\cite{ORourke}

In this study, we aimed to understand the binding modes, electronic structures 
and optical spectra of dye sensitizers on anatase TiO$_2$ nanowires having (101) 
and (001) facets by standard as well as hybrid density functional theory (DFT) 
calculations. Tetrahydroquinoline based C2-1 (C$_{21}$H$_{20}$N$_2$SO$_2$) 
organic dyes  achieving directional charge distribution upon photo-excitation 
have been considered as the light harvesting molecules adsorbed on the nanowires. 
In order to make comparisons with a simple dye molecule, we also included 
coumarin (C$_9$H$_6$O$_2$) which is extensively studied in the 
literature.\cite{Hara2,Wang4,Ehret,Wang5,Wang6,Wang7,Horiuchi,Kitamura,Wang8,Hagberg,Li,Chiba,Ito,Wang9,Kurashige,Zhang,Preat,Sanchez,Oprea} BecauseDSSC operates in solution, the solvation effects become important. Therefore, we addressed this by using a new 
polarizable continuum model (PCM) for solvents with different ionicities. 
After briefly describing the computational methods we will discuss the results 
in detail.

\begin{figure*}[t!]
\epsfig{file=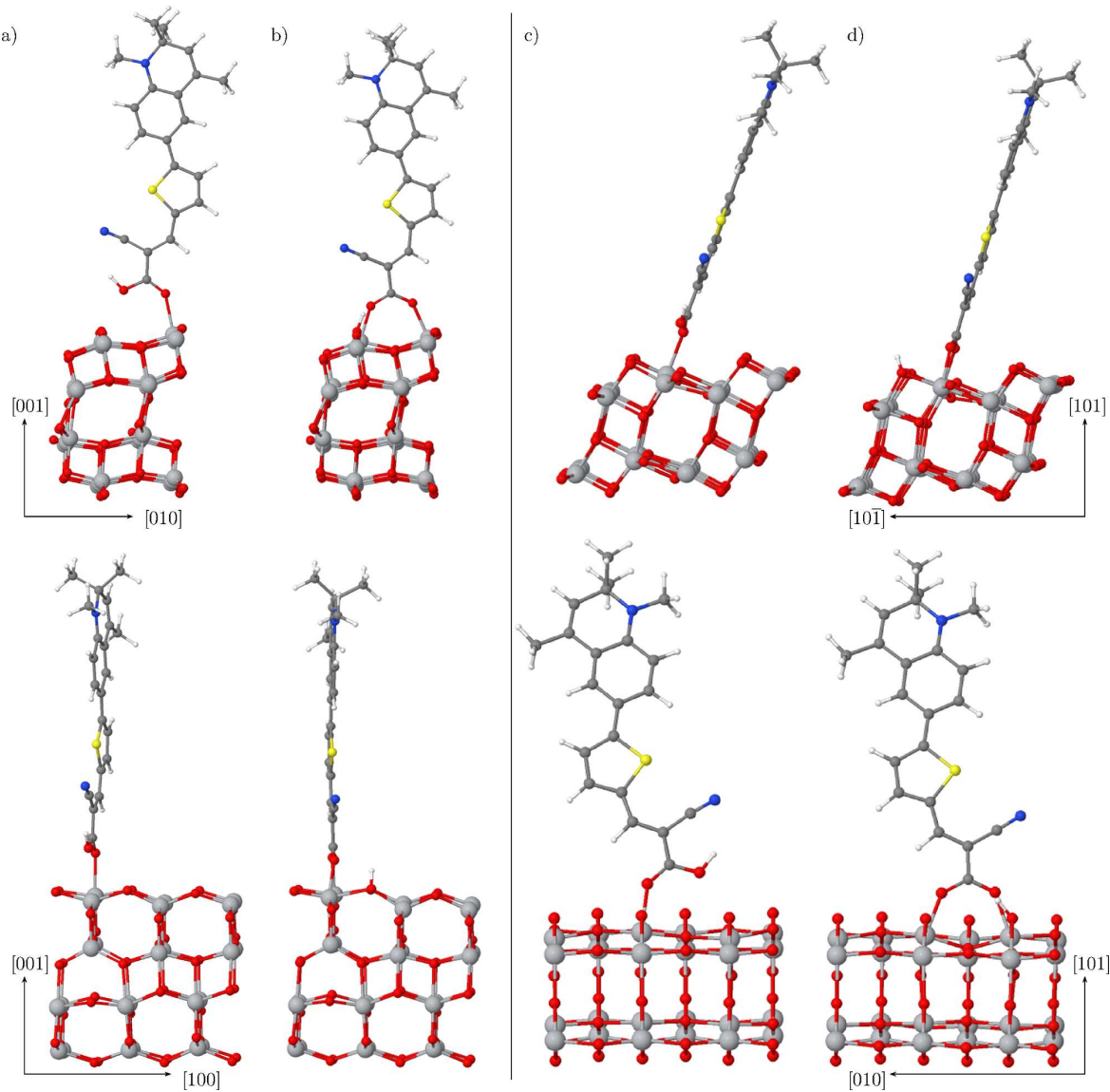,width=15.5cm}
\caption{Optimized adsorption geometries of C2-1 on anatase (001)-nanowire 
(left panel) and (101)-nanowire (right panel) with two different viewing 
orientations for both monodentate and bidentate bonding modes. Here the red, 
light-grey, dark-grey, blue, yellow and white color spheres represent O, 
Ti, C, N, S, and H atoms, respectively. \label{fig2}}
\end{figure*}

\section{Computational Details}

In order to investigate the geometric and electronic properties of the dye and 
nanowire composite systems, we performed pseudo potential plane wave 
calculations based on standard and hybrid DFT using both Perdew-Burke-Ernzerhof 
(PBE)~\cite{Perdew} and Heyd-Scuseria-Ernzerhof (HSE)~\cite{Heyd1,Heyd2,Paier} 
exchange-correlation (XC) functionals as implemented in the Vienna ab-initio 
simulation package (VASP).\cite{Kresse1} The ionic cores and valence electrons 
with an energy cutoff value of 400 eV for the plane wave expansion were treated 
by projector-augmented waves (PAW) method.\cite{Blochl,Kresse2} The convergence 
of our calculated values have been carefully tested with respect to $k$-point 
sampling. 

In addition to the standard DFT, we used the range separated hybrid density 
functional approach that partially admixes exact Fock exchange and PBE exchange 
energies. These type of hybrids offer a better description of localized $d$ 
states and improve the energy gap related features over the standard XC 
schemes. Range separated hybrid functionals tend to compensate the localization 
deficiency due to the lack of proper self-interaction cancellation between the 
Hartree and exchange terms of the pure DFT. Correct description of the adsorbate 
driven gap states around the band edges is critical for the estimation of the 
photovoltaic properties. In HSE functional, the exchange energy is separated into 
two parts as the long-range (LR) and the short-range (SR), as,
\[
E_{\tiny\textbf{X}}^{\scriptsize\textrm{HSE}}=
a E_{\tiny\textbf{X}} ^{\scriptsize\textrm{HF,SR}}(\omega)+
(1-a)E_{\tiny\textbf{X}} ^{\scriptsize\textrm{PBE,SR}}(\omega)+
E_{\tiny\textbf{X}} ^{\scriptsize\textrm{PBE,LR}}(\omega)
\]
where $a$ is the mixing coefficient~\cite{Perdew2} and $\omega$ is the range separation 
parameter.\cite{Heyd1,Heyd2,Paier}  In this approach, the correlation part of the energy 
is taken from standart PBE.\cite{Perdew} 

In order to study the effect of the solvent environment (chloroform and water) 
on the electronic structure of the dye+nw combined systems , we carried out
calculations via PCM including both the new non-linear and its linear  
counterpart as implemented in the open-source code 
JDFTx.\cite{KWeaver, Gunceler,universal_pcm,Sundararaman} 

In polarizable continuum models (PCM), a dielectric medium surrounding the solute 
molecule is generally used to reproduce the solvent environment. Therefore,
free energies can be computed without the need for explicit 
thermodynamic sampling of the many possible different configurations of solvent 
molecules. The surrounding cavity of the solute is modeled by its electron
density where the dielectric function of the solute turns on around a 
critical density value. In order to reproduce experimental solvation energies,
we parametrize the critical electron density, and the effective tension in the solute-solvent 
interface, accordingly.\cite{Gunceler,universal_pcm} The nonlinear\cite{Gunceler} PCMs are advantageous
over linear ones since the nonlinear models also incorporate the dielectric saturation effect.
In other words, the rotational contribution to the dielectric function decreases with 
increasing external field. For this, we separate out the rotational 
contributions from electronic/vibrational contributions, and model the former 
as a field of interacting dipoles. For a more detailed discussion of 
PCMs, we refer the reader to the respective reference.\cite{Gunceler}

\begin{figure*}[htb]
\epsfig{file=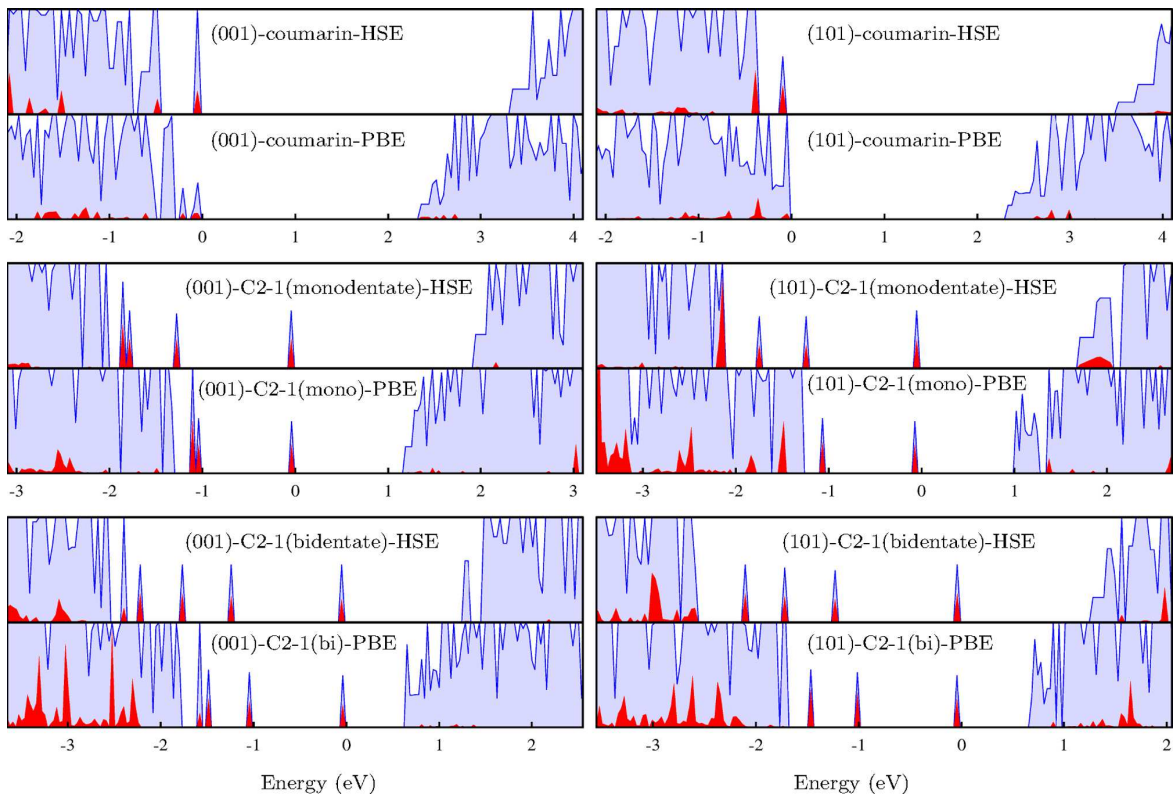,width=15.5cm}
\caption{Density of states (DOS) of dye-nanowire combined systems. The contribution of the dye molecules are denoted as red shades by projecting on the dye molecular orbitals. The zero of the energy is set at the highest occupied energy level.\label{fig3}}
\end{figure*}

\begin{figure*}[htb]
\epsfig{file=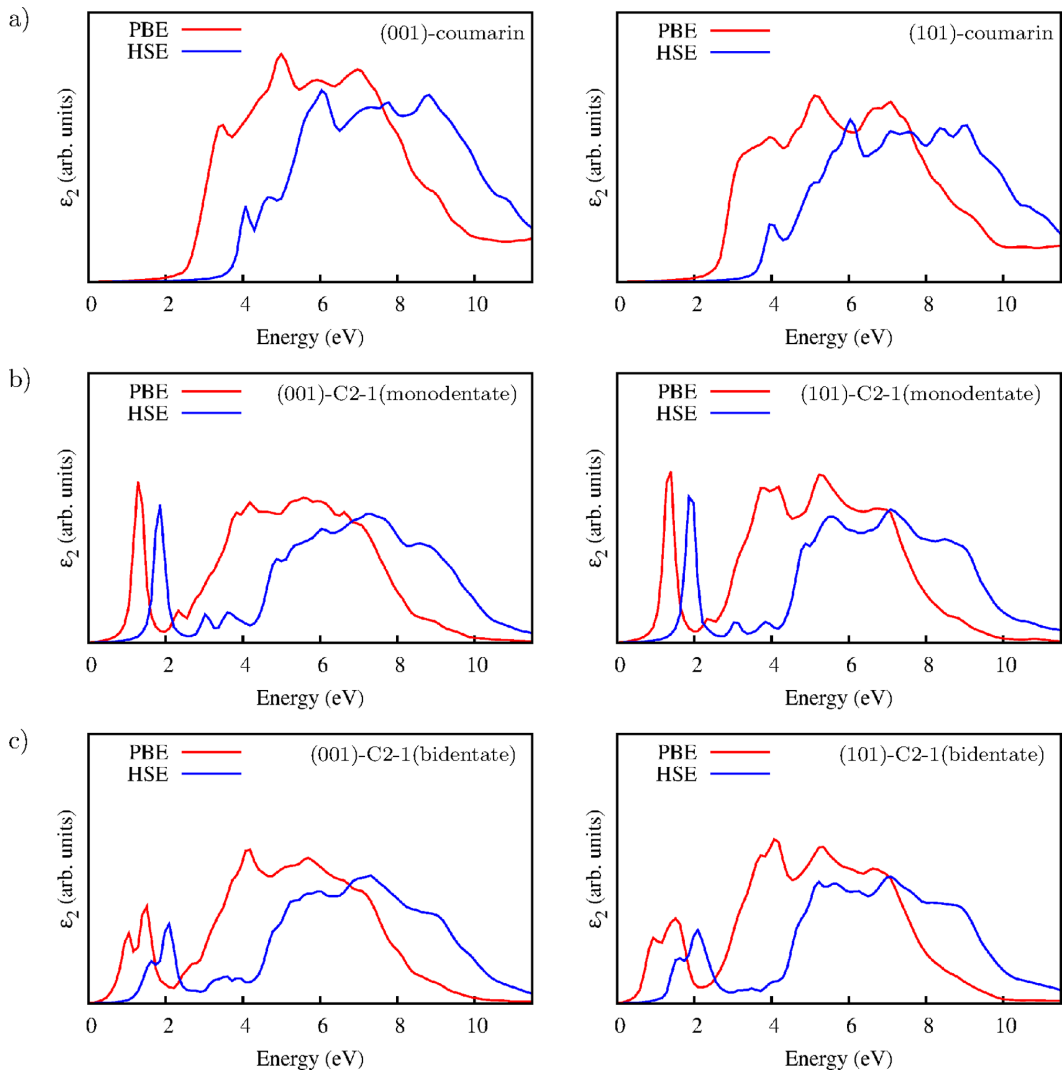,width=14cm}
\caption{Absorption Spectra for (a) Coumarin on anatase (001)-nanowire and (101) 
(b) C2-1 monodentate bridging on anatase (001)-nanowire and (101)-nanowire 
(c) C2-1 bidentate bridging on anatase (001)-nanowire and (101)-nanowire, respectively
\label{fig4}}
\end{figure*}

The stoichiometric nanowire models with (001) and (101) facets were built from 
the anatase form of bulk TiO$_2$,  referred as nw(001) and nw(101), 
respectively. Our tests show that the passivation of the facets are not 
required. Bare and dye adsorbed nanowire models are placed in periodic 
tetragonal super cells.  In order to prevent interaction with the periodic 
images of the structures a sufficiently large vacuum space of at least 20 {\AA} 
thick perpendicular to the nanowire axis has been inserted. Similarly, periodicity is 
enlarged along the nanowire axis such that the dye adsorbate can be assumed as 
isolated. The Hellman--Feynmann forces on each atom have been minimized 
($< $ 0.01 eV/{\AA}) based on the conjugate-gradients algorithm to fully 
optimize the initial geometries. The optimized nanowire models have maintained
the anatase structure as discussed in detail previously elsewhere.\cite{Unal}

\section{Results \& Discussion}

The coumarin core (C$_9$H$_6$O$_2$) has been focused on as a candidate sensitizer 
for creating highly efficient DSSCs by many experimental and theoretical 
works.\cite{Hara2,Wang4,Ehret,Wang5,Wang6,Wang7,Horiuchi,Kitamura,Wang8,Hagberg,Li,Chiba,Ito,Wang9,Kurashige,Zhang,Preat,Sanchez,Oprea}. 
First of all, the adsorption of coumarin core on the TiO$_2$ nanowires has been 
investigated as a minimal atomistic model to understand dye sensitization within 
the framework of total energy DFT calculations. We considered the coumarin 
molecule at various adsorption sites on the (001) and (101) facets of the anatase 
nanowires. The energetically favorable binding modes of coumarin for both of the 
cases are presented in Fig.~\ref{fig1}. Because of the tale oxygen of coumarin 
interacting with a five-fold coordinated surface Ti, the dye molecule aligns 
perpendicularly on both of the nanowire types forming a single bond (monodentate 
binding). Adsorption of coumarin does not cause noticeable distortion on the 
nanowire structure in both of the case. The length of the bond between the 
surface and the dye is 2.18 {\AA} and 2.21 {\AA} in the cases of nw(101) and 
nw(001), respectively.

In the case of C2-1 dyes, the tail oxygen and the OH group play an important 
role in the adsorption on the nanowire surfaces. The optimized geometries of 
different binding modes of C2-1 on TiO$_2$ nanowires are shown in 
Fig.~\ref{fig2}. It portrays two different adsorption modes. Similar to the 
coumarin case, one of them is the monodentate binding in which the tail oxygen 
forms a single bond with the surface Ti atom giving C2-1 a perpendicular 
orientation with respect to the nanowire axis. In the second one, in addition 
to the O-Ti bond, the OH group loses the H to the nearest surface oxygen site 
enabling the formation of a second O-Ti bond between the dye and the nanowire. 
This bidentate mode is a chemical binding and so is much stronger than the 
monodentate case. For both of the nanowire types, the monodentate bond length 
is slightly larger than bidentate formation where the bond lengths are $\sim$ 
2.0 {\AA} which is a typical value of the bulk structures. The adsorption of 
C2-1 only leads to a minor change in the local environment on the nw(001).
However,  the surface oxygen on the nw(101) in between the two Ti atoms forming 
the bidentate bonds slightly buckles down as seen in Fig.~\ref{fig2}d.

We calculated the binding energies with a standard formulation as given in 
previous studies.\cite{Cakir1,Cakir2} These values are computed using PBE and 
HSE schemes and are presented in Table~\ref{table1}.  The solvent effects
with chloroform and water solutions have also been included on PBE results. 
Single bonding gives moderate and similar binding energies for both of the 
dyes on the two nanowire surface types. Regarding the binding energies, HSE 
functional yields similar values compared to the PBE ones. Strong solution 
effects are found to be drastically weakening the adsorption of the dye on the 
nanowire surface. This could indicate an expectation of low efficiencies for 
singly bound molecular cases because of the reduced stability.  When the 
binding energies of the dyes on the nanowires in Table~\ref{table1} are 
considered, one can say that the dissolution of the simple coumarin dye in 
the electrolyte is probable in applications which causes noticeable decrease 
in the photovoltaic performance of DSSCs. However, the formation of the second 
bond as a result of the loss of the H in the OH group in the bidentate case of 
C2-1 dye enhances the binding appreciably. This bidentate mode of C2-1 gives 
the strongest binding energy among the cases considered here. Our 
PBE-calculated value of 0.94 eV on nw(101) is in good agreement with GGA 
predicted binding energy of the isolated adsorption of the molecule on (101) 
slab surface by O'Rourke \textit{et al.}\cite{ORourke} The slight buckling of 
the surface oxygen on nw(101) brings an energetic penalty on the binding energy 
of C2-1 bidentate mode which leads to a reduction of the adsorption energy 
compared to that of the nw(001) case. Our results suggest that C2-1 is losing 
the H from its tale OH group to the nanowire surface to form a chemical bond. 
Therefore, it survives in a strongly ionic solution like water without 
dissociation.  

The density of states (DOS) of the dye+nanowire combined systems have been 
presented in Fig.~\ref{fig3}. Our TiO$_2$ nanowire models having diameters 
around 1 nm possess larger band gap values relative to the bulk and surface 
structures as a result of the quantum confinement effect.\cite{Unal} In general, 
most of the deep lying occupied molecular orbitals stay in the valence band as 
a resonant state. Most importantly, a number of dye-related isolated and occupied 
states appear above the VB edge within the band gap, depending on the nature of 
the binding. As a result, the Fermi energy shifts up to higher energies leading 
to a energy gap narrowing which is an important factor for photovoltaic 
properties. Moreover, the lowest lying unoccupied molecular levels of the dyes 
delocalizes on the Ti 3$d$ states inside the conduction band (CB) of the 
nanowires.  HSE functional corrects the band gap underestimation which is 
inherent in the standard PBE XC scheme. The PBE-bandgaps of bare nw(101) and 
nw(001) are 2.51 eV and 2.69 eV, respectively. HSE method gives an energy 
correction of 1.50 eV for nw(101) and 1.37 eV for the nw(001). This opening of 
the gap results in molecular states to fall in the energy gap, which were 
previously to be in resonance with the VB at the PBE level. In order to compare 
PBE and HSE calculated DOS structures, we aligned them with respect to their 
deep core energy states. For both of the nanowires, the molecular states of 
coumarin appear around the VB edge while one of them is isolated from the rest 
in the case of HSE. For C2-1 monodentate mode, PBE predicts two isolated states 
above the VB. Due to the gap opening by HSE functional, one more filled 
isolated state falls in the band gap of both of the nanowire types. On going 
from monodentate to bidentate bonding, there appears an additional isolated 
state above the VB. In the case of C2-1 bidentate mode, the positions and the 
number of dye related energy levels calculated with the PBE functional are in 
agreement with the GGA results of O'Rourke \textit{et al.}\cite{ORourke} The 
only difference comes from the alignment of TiO$_2$ energy bands with respect 
to the energy levels of the dye, which stems from our nanowire and their slab 
calculations. Significant band gap reduction is obtained in the case of C2-1 on 
both of the nanowires which is important for light harvesting.

The absorption spectra of the sensitizers, coumarin and C2-1, on nw(001) and 
nw(101) have been calculated at the PBE and HSE levels using the formulation as 
described in our previously study on the bare anatase nanowires.\cite{Unal} The 
dye related contributions in the optical spectra show similar characteristics 
with both PBE and HSE calculations as shown in Fig.~\ref{fig3}. These features 
correspond to the first absorption peaks in each case associated with 
excitations from the highest occupied molecular states to the unoccupied 
molecular states which are coupled to the CB of the semiconductor. Similar 
absorption properties are obtained in the UV region, which are mostly related 
with the interband transitions from the VB to the CB states. Although the main 
characteristics are alike, PBE-calculated spectra is significantly red shifted 
with respect to that of the HSE due to the local density approximation (LDA) 
giving rise to an inherent underestimation of the band gap of TiO$_2$. The 
coumarin core brings the lowest lying peak which extends the absorption 
threshold slightly into the visible region as seen in Fig.~\ref{fig4}a for 
both of the nanowire types. The C2-1+nanowire combined system has more 
favorable optical properties than the coumarin+nanowire structures. A distinct 
and strong absorption peak results due to the excitation from the occupied 
molecular state at the Fermi energy to the frontier unoccupied molecular state 
which delocalizes inside the CB of the semiconductor.  As being a D-$\pi$-A 
type dye,  such an excitation achieves charge redistribution from the donor to 
the acceptor moiety of C2-1 dye which is important for the charge injection 
into the CB of TiO$_2$.  This peak associated with the charge transfer (CT) 
state is identified at $\sim$2 eV at the HSE level while PBE prediction 
falls in the near infrared (IR) part of the spectrum. In the energetically 
the most preferable bidentate bonding mode of C2-1 sensitizer, two absorption 
peaks are identified which are significantly broadened in the visible range. 
This is reminiscent of the chemical nature of the adsorption of C2-1 on both 
the (001) and the (101) surfaces of the anatase nanowire. The first absorption 
peak positions of C2-1 on the anatase nanowire structures calculated at the HSE 
level are slightly red shifted compared with respect to the experimental 
data\cite{Chen1,Chen2} of Chen \textit{et al.} obtained on nanocrystalline 
TiO$_2$. Consequently, based on our hybrid DFT computations, the bidentate 
bonding form of C2-1 on TiO$_2$ survives in solution,  causes a number of 
isolated filled molecular states to appear in the band gap, functionalizes the 
anatase nanowires to absorb a wide range of the visible region and achieves 
charge separation which is promising for both the enhancement of the charge 
injection efficiencies and for the reduction of the charge carrier recombination 
rates.

\section{Conclusions}

The binding geometries, electronic structures and  absorption characteristics of 
two organic molecules on TiO$_2$ nanowires with (101) and (001) facets have 
been investigated using screened Coulomb hybrid density functional calculations. 
Coumarin causes new gap states to appear at edge of the VB of  the nanowires. 
These systems that we chose as the minimal dye+nanowire models result in a 
narrowing of the electronic band gap of TiO$_2$. Strong bidentate binding of 
tetrahydroquinoline C2-1 dye brings a number of isolated and occupied gap states 
that both result in a significant narrowing of the band gap and cause a broader 
absorption structure functionalizing the semiconductor nanowires in whole visible 
region. Nonlinear solvent effects suggest that the dissolution of coumarin and 
monodentate binding of C2-1 from the nanowires is probable in an actual 
electrolyte. Bidentate adsorption of donor-$\pi$-acceptor type C2-1 molecules 
can achieve directional charge transfer excitation to increase charge injection 
probabilities, allow absorption in the full range of visible spectrum to achieve  
enhanced light harvesting, and exhibit strong binding to reduce degradation of 
possible device operation.

\begin{acknowledgments}
This work is supported by T\"{U}B\.{I}TAK, The Scientific and Technological
Research Council of Turkey (Grant \#110T394). Computational resources were
provided by ULAKB\.{I}M, Turkish Academic Network \& Information Center.
\end{acknowledgments}

\end{document}